\documentclass[12pt]{article}
\usepackage{mathtext}
\usepackage{graphicx}
\usepackage[english]{babel}
\usepackage{amsfonts}
\usepackage{amssymb}
\usepackage{amsmath}
\newcommand{\ar}{\longrightarrow}

\newcommand{\la}{\lambda}

\begin{document}
\title{Quantum parallelism may be limited}

\author{Yu.I.Ozhigov\thanks{ozhigov@cs.msu.su}\\
Moscow State University of M.V.Lomonosov, VMK Fakulty,\\
Institute of physics and technology RAS}
\maketitle
PACS: 03.65,  87.10\\

\begin{abstract}

We consider quantum formalism limited by the classical simulating computer with the fixed memory. The memory is redistributed in the course of modeling by the variation of the set of classical states and the accuracy of the representation of amplitudes. This computational description completely preserves the conventional formalism and does not contradicts to experiments, but it makes impossible fast quantum algorithms. This description involves the slow down of quantum evolutions with the growth of the dimension of the minimal subspace containing entangled states, which arise in the evolution. This slow down is the single difference of the proposed formalism from the standard one; it is negligible for the systems from the usual experiments, including those in which many entangled particle participate, but grows rapidly in the attempt to realize the scalable quantum computations, which require the unlimited parallelism. The experimental verification of this version of quantum formalism is reduced to the fixation of this slow down. 
\end{abstract}
\section{Introduction and background}

The algebraic apparatus of quantum theory is marvellous in its simplicity. The time evolution $|\Psi(t)\rangle$ of the state vector of any system in the Hilbert space  ${\cal H}$ is described by the unitary operator $U_t$ of the form $U_t=exp(-\frac{i}{h}Ht)$, where $H$ is Hamiltonian, and the exponential is generally speaking chronological. The state in the moment $t$ is given by $|\Psi(t)\rangle=U_t|\Psi(0)\rangle$.

This formalism is very simple from the algebraic view point but its computational cost is exponential. Its drawback does not manifest itself when the evolution can be found symbolically, e.g. "with a pencil in a hand"  that is the source of the success of quantum theory, because for the model problems, even with the huge formal dimensionality of Hilbert space ${\cal}H$ this is the case. But the simulation of complex systems going far beoynd standard tasks requires the usage of computers, and here the algebraic beauty turns into an unpleasant side: exponentially growing complexity. 

Feynman's idea: to build quantum computer as the simulation of a real system by the set of qubits with the size of the order of the initial system was designed to overcome this curse of dimensionality (the exact formulation and the proof can be found in the work \cite{Za}). The realizability of this plan is doubtless. But its realization by a quantum gate array proposed by Feynman requires the more profound analysis of the many body quantum formalism than it has been done up to now. 

Quantum algorithm is the recipe of the classical change of Hamiltonian that is the evolution $H(t)$, such that the corresponding evolution of the state vestor $|\Psi(t)\rangle$ gives the desired distribution of amplitudes. In the case of the modeling of a real system the desired is the discret approximation of the real state vector, which we cannot control straighforwardly unlike quantum computer. There are another tasks, purely mathematical, for which the application of quantum computer promises not only the convenience but the direct gain in the time in comparison with any possible classical computer. This is the problem of search in the different formulations, in general term: the search of the solution of the equation  $f(x)=1$ for the Boolean function $f$ from $n$ variables (see \cite{Sh},\cite{Gr}). Beyond the evident practical importance the search problem gives the single criterion of that we have built just quantum computer and not its imitation. 

Experiments been held from the end of 80-th up to now do not show the possibility to make combinations of quantum gates scalable up to the valid quantum computer, which could either outperform calssical counterparts, or at least demonstrate such ability in a foreseeable future. 
 In all likelyhood, quantum theory of many bodies requires the more profound elaboration. It concerns the complexity of quantum evolutions when the dimensionality of Hilbert space of states becomes too large to make possible to handle with it effectively. 

In this work we show the simple description of many body evolutions based on classical algorithms with the flexible redistribution of memory. It is effective from the view point of classical complexity theory that allows the experimental check and practical simulation on quantum level as well. This method preserves the possibility to build every states in Hilbert space, and in this sense it gives nothing different from the standard theory. However, the cost of the computational effectiveness is the direct ban on fast quantum algorithms. 

This method is to a certain extent modification of quantum theory, but it touches only the area where this theory has been never checked - Hilbert spaces of inaccessible large dimensionality. This must not be confused with the large number of particles as in quantum ensembles treated in quantum theory of many bodies. For example, for the vibrations of solid state, in which billions of atoms participate in entangled states but the dimensionality of Hilbert space, however, is small. 

Computer simulation of the reality is an algorithm (classical), which for every pair: initial state + tme moment, gives the approximation of the final state of the considered system in this moment. If a state is treated as the wave function, the simulation acts as $(|\Psi(0)\rangle,\ t)\ar |\Psi(t)\rangle$. The simulation is adequate if the time complexity of this algorithm is $O(t)$, e.g. differs from the real time only in the coupling coefficient. We accept the thesis of algorithmic physics, or constructivism, that is any real process has the adequate simulation.  This is methaphysical thesis. It can be reformulated so that the physical time is measured by some abstract algorithm, which number of steps has thus the sense of time. The presence of moment of the measurement in the incoming data is important because we do not require that the algorithm works in a real time mode, e.g. would be ready to give the answer $|\Psi(t_0)\rangle$ in arbitrary moment $t_0$  when we decide to interrupt its work. The observation of a real system includes the preliminary knowledge about the time in which we intend to do it. 

The important property of the simulation algorithm is the limited memory, which can be flexibly redistributed in course of computation. The set of classical states is not thus fixed (as in quantum informatics) and can vary depending on the conditions of the real system. The representation of a quantum state, pure or mixed, arisen in the evolution leads to the reduction of the classical states number so that in any case the total memory is fixed. 

The sense of constructivism is that we can control all the sides of real processes but the pure probability, which appears in quantum measurements. Constructivism permits the existence of a quantum computer as a physical model of the complex process in the case when the building of its computer simulation is hard, for example, if the coupling coefficient is large. At the same time constructivism forbids fast quantum algorithms e.g. processes, which we cannot reproduce via classical algorithms of the complexity, which grows as the real physical time. Hence constructivism can be refuted experimentally; for this one must demonstrate really working quantum speedup of the search problem over classical computation that seems to be hardy realizible task in view of the modern state of experiments. 

The practical scalable realization of the fast quantum computation is the single way to the refutation of the proposed computational correction to quantum formalism. No other experiments can do this, because all conclusions made in any physical theories (including quantum), which were checked in experiments up to now were made in the paradigm of classical computations. 

\section{Computational "relativism" of quantum theory}

Let Hilbert space be fixed. How looks the realization of the operator $U_t$ on a classical computer? We must sequentially take all basic states of Hilbert space ${\cal H}$, at each step applying $U_t$ to the current state. The postulate of quantum parallelism claims that all basic states are processed simultaneously and just this postulate makes possible fast quantum computations. 

But if this postulate is wrong, we have to suppose that basic states in course of application of $U_t$ are processed successively, one after another. If the time frame $t$ in which the evolution operator $U_t$ acts, is small enough (in the proof of equivalence between path integrals and Shredinger equation it is required that $t=O(\delta x)^2$ where $\delta x$ is the value of spatial resolution), $U_t$ is close to the identity operator  $I$, and the successive processing of the basic states gives the same resut as the parallel, only with the slow down proportional to the dimensionality of ${\cal H}$. 

The time of the work of the optimal classical algorithm must coincide to the real physical time $t_{phys}$ within the coefficient. This thesis requires the exact definition of what is the optimal algorithm. We rest on the definition of complexity as the number of elementary operations as in algorithm theory. The total memory is fixed and we have to redistribute it between the virtual devices corresponding to global parameters of the model: the number and the form of classical states, the accuracy of quantum amplitudes representation. 

Such redistribution is not fixed beforehand and is determined by the condition of experiment. There is nothing different in this indeterminicity  from the physical experiments, which require the preliminary tuning of devices. The framework of optimal tuning is not exact. However, it does not interfere the experimental check of the proposed correction to formalism. The point is that the difference between the rival computational models is not so valuable than the large difference between the standard time $t$ and the "computational" time  $t_{phys}$ for the large dimensionality of Hilbert space. This difference grows rapidly with the complication of the simulated system that gives hope to observe it almost independently of how close to ideal is the used simulating algorithm.

In this work we do not touch the question of the flexible memory redistribution. In what follows we consider for simplicity the case of fixed global parameters. 

We represent all classical states in the form of points in the digital  $n$- dimensional lattice $Z_n=Z/nZ$, which nodes we call cells. Let each cell $|j\rangle$ contain the complex number $\la_j$, equal the amplitude of wave function on this basic state. Let Hamiltonian of the system have the form $H_{ord}\sum\limits_j\frac{p_j^2}{2m_j}+\sum\limits_{i,j}V_{i,j}$ where indices take the values of numbers of real particles. The action of operator $U_t$ has the form of the work of Turing machine on the tape of the form $Z_n$, which has one head that can observe $2n+1$ cells at the same time: one is the main and  $2n$ neighboring. Each step of its work is the application of the scheme of finite differences realizing the energy operator. The movements of the head are so that it successfully comes through each cell in the framework of the considered array exactly once. This is the successful method of the application of $U_t$ for a given Hamiltonian.  

We note that due to the smallness of $t$ the concrete arrangement method for ordering of the nodes of $Z_n$ plays no role, the only important is that the head passes through each cell only once. 

The time of the work of this Turing machine is $N\delta t$, where $N=dim({\cal H})$ is the number of all cells, $\delta t$ is the time, required to fulfil the scheme of finite differences. The last value though it grows with $N$, it is however very small compared with $N$, hence the main deposit to the time gives just $N$. When the number of particles $N$ grows exponentially, the slow down in the seccessful realization of evolution will be exponential as well. 

If Hamiltonian has another form, for example, it can be obtained from $H_{ord}$ by the change of basis: $H=T^{-1}H_{ord}T$, the evolution $U_t=exp(-\frac{i}{h}Ht)$ is obtained from $U_{t\ ord}=exp(-\frac{i}{h}H_{ord}t)$ by the same change: $U_t=T^{-1}U_{t\ ord}T$. If $H_{ord}$ is the sum of members, each of which depends on only variables, corresponding to one particle, the complexity of the simulation of such evolution depends on the complexity of $T$ but not on $t$. 

For example, for a system of interacting harmonic oscillators with coordinates $\bar x$ we can pass to quasi particles - phonons with coordinates $\bar X$ by the canonic transform. The corresponding matrix $T$ has exactly one unit in each row and in each column, e.g., it is the permutation of basic vectors. 

Indeed, the passage to new phonon coordinates has the form  $\bar x=F(\bar X )$, where $F$ is Fourier transform on classical coordinates. In Hilbert space of quantum states this transform has the form "point to point", that is the permutation of basic vectors, like CNOT. $T$ will be as CNOT disentangling operator, and phonons will not be entangled; this is the sense of quasipartices. Simulation of such a system is reduced to the operator $T$, which does not depend on time. Here we do not touch the question about the accuracy on the classical space, which determins coupling constant. 

Let us now consider the work of Grover search algorithm (GSA) in the parallel (as in standard formalism) and in the successive realization of quantum formalism. Initially amplitudes of all states  $|j\rangle$ are equal to  $N^{-1/2}$. In the process of application of $U_t$ - its role plays operator $G$, the small portion of amplitude is taken from each state, and (after small correction, connected with unitarity) is added to the amplitude of state $|j_{target}\rangle$, which corresponds to the solution of equation $f(x)=1$. The target state gradually gathers amplitude from all states to itself. The speedup in comparison to classical search is that this quantum process goes in parallel mode.

In the successful processing of the state vector we must look over all $N=2^n$ states one after another and we have no gain despite that all states in GSA will be the same, only they alternate much slower. 

\section{Pilot}

Constructivism presumes the peculiar building of algorithms when the memory is redistributed flexibly; to realize it we need the specific programming. We briefly describe here the possible approach to this programming; the further elaboration of it is the topic of the more detailed work. 

We call a pilot Turing machine, which it performs the computation of the wave function (see above). This term is taken from de Brohile pilot wave, which idea is the similar. A pilot is a theoretical object and no physical sense can be ascribed to it but the aim to which it is designed: computation of the wave vector evolution.    If the system of two particles is in nonentangled state and the particles do not interact with each other it is naturally to assume that each of them has its own pilot and their evolution goes independently. As they interact and become entangled one common pilot arises instead of two. The same is true for $n$ particles. The inverse process: replication of pilots takes place in the partial measurements of wave vector. We treat that all states, which are not observed by the pilot head in some moment are the subject of Lindblad operators that leads to the mixed state described by the density matrix.  

To extend directly our scheme to density matrices woud be needed to introduce the classical space of pairs $Z_{2n}=Z_n\times Z_n$, so that the head of Turing machine observes two classical states simultaneously. But it would not be right because in the reality the decoherence leads to the partial measurement and gives the certain pure states only we do not know which. The evolution of such state must go with the speed independent from the other states in the mixture. Whereas the "coupled" Turing machine speed depends on them that is not right. Hence we must consider the mode of evolution as branching after the decoherence operators action so that all components of the mixture are processed independently like the independent particles. Here in the components corresponding to the measuremet will be non entangled parts, which of them must have its own pilot. This is the replication of pilots in the decoherence. The accuracy of representation of the wave function: total number of classical states and the accuracy of amplitudes  decreases in such replication. 

A polit may be designed as follows. Let we have an EPR pair $\frac{1}{\sqrt 2}(|01\rangle+|10\rangle )$, we apply to the first qubit the operator $U_1$, whereas the second qubit is inaccessible, for example, it is located at far distance. Instead of keeping in memory all states of 4 components we can do the following. We will store the amplitudes of states $U_1|0\rangle_1$ and $U_1|1\rangle_1$, and add to the states of the first qubit $|0\rangle_1$ and $|1\rangle_1$ the labels, which point that their remote component will be  $|0\rangle_1$ and $|1\rangle_2$ correspondingly. Further we can apply operators to the second qubit analogously. Any change on them will have no influence on the first qubit. We consider the cost of labels as negligible compared to the real operations. This cost corresponds to the spatial separation of qubits. The computational approach thus does not permit to overcome the relativistic limitation on the speed of information transfer that is the important property of quantum formalism. 

If the real system consists of many components $S_1,S_2,\ldots, S_n$ and we apply some operator on only one subsystem, for example, on $S_1$, then on $S_2$ etc., and then redistribute particles on subsystems, we obtain the entangled states, and in the further modeling we have to apply the labels for each of subsystems. Here the practical problem of the modeling of entanglement dynamics on a classical computer arises that presumes the flexible distribution of the memory. For the fixed global parameters this description is straightforward.
This description is done in the standard formulation of quantum informatics reduced to linear algebra. The problem of redistribution the parameters for the optimal model ready for the comparison with experiments is the special topic, which development promises new perspectives in the theory of quantum computer. 

We also note that the pilot representation touches the notion of elementary events, which form quantum probability. It is naturally to assume that the measurement of quantum state gives the basic state, which the pilot observes at the moment. The movements of Turing machine head during the optimal simulation must be arranged so that the time spent at some area of basic states should be proportional to the quantum probability of this area. The supposition that the result of the measurement coincides to the state observed by the pilot thus gives just quantum probability. 

\section{Conclusion}

We have considered constructive thesis about the absolute priority of classical algorithms in the description of real evolutions. This thesis limits quantum parallelism by some coupling constant between quantum dynamics and the size of real ensembles. This limitation is equivalent to some slow down of real time  $t_{real}$ in comparison with the abstract time $t$ from Shredinger equation: $U_t=exp(-\frac{i}{h}Ht)$. The slow down is very small for cases when the classical computation simulates quantum evolution adequately but it becomes very large if the fastest classical computation begins to lag behind the abstract time $t$. It reveals in fast quantum algorithms, which turn to be impossible in the proposed computational formalism just because such algorithms require exponentially large virtual resource, which already for the modest size of memory begins to exceed all real ensembles. 

This hypothesis can be refuted by the realization of fast quantum computations. The confirmation of the hypothesis we can obtain if the time slow down will be discovered. The last would be possible in the attempt to realize GSA with the good suppressing of decoherence; this may be accessible with quantum codes of correction (see, for example, \cite{Sh2}). Spreading of quantum state on the large set of real particles can support the conservation of it in the time and makes possible its slow evolution when the real physical time  $t_{phys}$ will exceed exponentially the formal time $t$ from the standard formalism. 

The important step in the confirmation of computational hypothesis is the programming of new type, in which the fixed memory of simulating computer is distributed most flexibly. The number of classical states, their form and the accuracy of the amlitude representation are not thus fixed magnitudes as in the simulation based on the standard treatment of Turing machine or the other models of computation. The time of such hybrid computation must give within a coefficient the exact value of physical time $t_{phys}$, which difference from the formal $t$ an experiment must fix. The creation of such computing model is the separate mathematical problem. 

Experimental discovery of the difference between $t$ and $t_{phys}$ in the case of good suppression of decoherence would have the big value because it would open door for the effective simulation of complex systems using the whole capabilities of quantum theory. The sacrifice in the form of fast quantum algorithms, which would be made for this seems to be fully justified.

\end{document}